\documentclass[manuscript]{aastex}
\usepackage{spr-astr-addons}

\usepackage{hyperref}
\usepackage{multirow}
\usepackage{multicol}

\shorttitle{On the limit between short and long GRBs}
\shortauthors{Tarnopolski}

\begin{document}

\title{On the limit between short and long GRBs}

\author{M. Tarnopolski}
\affil{Astronomical Observatory of the Jagiellonian University\\ul. Orla 171, 30-244 Krak\'ow, Poland}
\email{mariusz.tarnopolski@uj.edu.pl}

\begin{abstract}
Two classes of GRBs have been identified thus far without doubt and are prescribed to different physical scenarios -- NS-NS or NS-BH mergers, and collapse of massive stars, for short and long GRBs, respectively. The existence of two distinct populations was inferred through a bimodal distribution of the observed durations $T_{90}$, and the commonly applied $2\,{\rm s}$ limit between short and long GRBs was obtained by fitting a parabola between the two peaks in binned data from BATSE 1B. Herein, by means of a maximum likelihood (ML) method a mixture of two Gaussians is fitted to the datasets from BATSE, {\it Swift}, {\it BeppoSAX}, and {\it Fermi} in search for a local minimum that might serve as a new, more proper, limit for the two GRB classes. It is found that {\it Swift} and {\it BeppoSAX} distributions are unimodal, hence no local minimum is present, {\it Fermi} is consistent with the conventional limit, whereas  BATSE gives the limit significantly longer (equal to $3.38\pm 0.27\,{\rm s}$) than $2\,{\rm s}$. These new values change the fractions of short and long GRBs in the samples examined, and imply that the observed $T_{90}$ durations are detector dependent, hence no universal limiting value may be applied to all satellites due to their different instrument specifications. Because of this, and due to the strong overlap of the two-Gaussian components, the straightforward association of short GRBs to mergers and long ones to collapsars is ambiguous.
\end{abstract}


	\section{Introduction}\label{intro}

Gamma-ray bursts (GRBs) were detected by military satellites \textit{Vela} in late 1960's. GRBs were recognized early to be of extrasolar origin \citep{klebesadel}. \citet{mazets} first observed a bimodal distribution of $T_{90}$ (time during which 90\% of the burst's fluence is accumulated) drawn for 143 events detected in the KONUS experiment. Burst and Transient Source Explorer (BATSE) onboard the Compton Gamma Ray Observatory (\textit{CGRO}) \citep{meegan92} allowed to confirm the hypothesis that GRBs are of extragalactic origin due to isotropic angular distribution in the sky combined with the fact that they exhibited an intensity distribution that deviated strongly from the $-3/2$ power law \citep{briggs,fishman}. However, a more complete sample of BATSE short GRBs were shown to be distributed anisotropically \citep{meszaros2,vavrek} and cosmological consequences were discussed lately \citep{meszaros}. BATSE 1B data release was followed by further investigation of the $T_{90}$ distribution \citep{kouve} that lead to establishing the common classification of GRBs into short ($T_{90}<2\,{\rm s}$) and long ($T_{90}>2\,{\rm s}$). This $2\,{\rm s}$ limit was derived by fitting a parabola to the local minimum of the binned distribution of 222 GRBs. It was observed that durations $T_{90}$ seem to exhibit log-normal distributions which were fitted to short and long GRBs \citep{mcbreen}, resulting in mean durations equal to $0.37\,{\rm s}$ and $26.36\,{\rm s}$. A mixture of Gaussians fitted to $\log T_{90}$ dataset from BATSE 2B yielded locations of the components equal to $0.60\,{\rm s}$ and $32.1\,{\rm s}$ \citep{koshut}, while a subset of BATSE 3B sample yielded $0.42\,{\rm s}$ and $34.4\,{\rm s}$ \citep{kouve2}. A complete BATSE dataset gave mean locations of the groups at $0.78\,{\rm s}$ and $34.7\,{\rm s}$ \citep{horvath02}. The progenitors of long GRBs are associated with supernovae \citep{woosley} related with collapse of massive, e.g. Wolf-Rayet, stars. Progenitors of short GRBs are thought to be NS-NS or NS-BH mergers \citep{nakar}, and no connection between short GRBs and supernovae has been proven \citep{zhang5}.

The existence of an intermediate-duration GRB class, consisting of GRBs with $T_{90}$ in the range $2-10\,{\rm s}$, was put forward \citep{horvath98,mukh} based on the analysis of BATSE 3B data. It was supported \citep{horvath02} with the use of the complete BATSE dataset. Evidence for a third normal component was also found in {\it Swift} data \citep{horvath08,zhang2,huja,horvath10}. {\it BeppoSAX} dataset was shown to be in agreement with earlier results regarding the bimodal distribution, and the detection of an intermediate-duration component was established on a lower, compared to BATSE and {\it Swift}, significance level due to a less populate sample \citep{horvath09}. It is important to note that in {\it BeppoSAX} only the intermediate and long GRBs were detected, the short ones being not present. Interestingly, \citet{zitouni} re-examined the BATSE current catalog as well as the {\it Swift} dataset, and found that a mixture of three Gaussians fits the {\it Swift} data better than a two-Gaussian, while in the BATSE case statistical tests did not support the presence of a third component. Regarding {\it Fermi}, a three-Gaussian is a better fit than a two-Gaussian\footnote{Adding parameters to a model always results in a better fit (in the sense of a lower $\chi^2$ or a higher maximum log-likelihood) due to more freedom given to the model to follow the data. The important question is whether this improvement is statistically significant, and whether the model is an appropriate one. See \citep{Tarnopolski,Tarnopolski2} for a discussion.}, however the presence of a third group in the $T_{90}$ distribution was found to be unlikely \citep{Tarnopolski,Tarnopolski2}.

The $2\,{\rm s}$ limit is widely used in GRB analysis. However, the {\it Swift} data were re-examined \citep{bromberg} and it was found that a limit of $0.8\,{\rm s}$ is more suitable for the GRBs observed by {\it Swift}. Many works in which a two-Gaussian was fitted to the $\log T_{90}$ distribution showed a significant overlap of components corresponding to short and long GRBs \citep{mcbreen,koshut,horvath02,zhang2,huja,horvath09,barnacka,zitouni}, regarding datasets from BATSE, {\it Swift}, {\it BeppoSAX}, {\it Fermi}, among others. The mentioned datasets consist of $\sim 1000-2000$ events. Based on the well-established conjecture that durations $T_{90}$ are log-normally distributed, the limit between short and long GRBs may be placed at the position of the local minimum of a mixture distribution.

The aim of this paper is to examine what limits are most suitable for GRB samples observed by different satellites. In Section~\ref{sect2}, the datasets and methods are described. Results are shown in Section~\ref{sect3}, while Section~\ref{sect4} is devoted to discussion and is followed by concluding remarks gathered in Section~\ref{sect5}.

\section{Datasets and methods}\label{sect2}

The datasets\footnote{All accessed on April 29, 2015.} from BATSE\footnote{\url{http://gammaray.msfc.nasa.gov/batse/grb/catalog/current}}, {\it Swift}\footnote{\url{http://swift.gsfc.nasa.gov/archive/grb table}}, {\it BeppoSAX}\footnote{\url{https://heasarc.gsfc.nasa.gov/docs/sax/sax.html}}, and {\it Fermi}\footnote{\url{http://heasarc.gsfc.nasa.gov/W3Browse/fermi/fermigbrst.html}} are considered herein. They contain 2041 GRBs (BATSE current catalog), 914 ({\it Swift}), 1003 ({\it BeppoSAX}), and 1596 ({\it Fermi}). Additionally, a subset of BATSE data, i.e. complete BATSE 1B sample containing 226 GRBs, is examined to compare with results of \citet{kouve}. For display purposes, histograms are plotted using the Knuth rule for bin width. Up to date, to the best of the author's knowledge, only \citet{horvath12} and \citet{qin} conducted research on a {\it Fermi} subsample, consisting of 425 GRBs from the first release of the catalog.

The fittings are performed using the maximum likelihood (ML) method \citep{kendall}. Having a distribution with a probability density function (PDF) given by $f=f(x;\theta)$ (possibly a mixture), where $\theta=\{\theta_i\}_{i=1}^p$ is a set of $p$ parameters, the log-likelihood function is defined as
\begin{equation}
\mathcal{L}=\sum\limits_{i=1}^N\log f(x_i;\theta),
\label{eq1}
\end{equation}
where $\{x_i\}_{i=1}^N$ are the datapoints from the sample to which a distribution is fitted. The fitting is performed by searching a set of parameters $\theta$ for which the log-likelihood $\mathcal{L}$ is maximized. The fitted function in this case is a mixture of two standard Gaussians, $\mathcal{N}(\mu,\sigma^2)$:
\begin{equation}
f_k(x)=\sum\limits_{i=1}^k A_i\varphi\left(\frac{x-\mu_i}{\sigma_i}\right)=\sum\limits_{i=1}^k \frac{A_i}{\sqrt{2\pi}\sigma_i}\exp\left(-\frac{(x-\mu_i)^2}{2\sigma_i^2}\right).
\label{eq2}
\end{equation}
Here, $k=2$, so the distribution is described by $p=5$ parameters: two means $\mu_1$, $\mu_2$, two dispersions $\sigma_1$, $\sigma_2$, and one weight $A_1$. The second weight is $A_2=1-A_1$ due to normalization. Normal distribution's PDF is denoted by $\varphi$.

To estimate the parameter errors, $\delta\theta$, a simple Monte Carlo technique called a parametric bootstrap \citep{efron1,efron2,efron3} is performed, i.e., having a distribution from Eq.~\ref{eq2} fitted, it is randomly sampled to create a set of $N$ random variates ($N$ being the same as in the original dataset). This set is used to find another fit. After repeating this procedure 1000 times, the standard deviations are computed from the 1000 sets of 5 parameters, and serve as errors for the parameters obtained from the original dataset.

In the same manner the error of the location of the local minimum is estimated. In this case, some of the realisations drawn from the original bimodal distribution may happen to be unimodal, hence no local minimum might be present. In that case, only a fraction of the 1000 realisations which do have a local minimum is taken into account. The opposite situation may also occur, i.e. sampling a unimodal distribution and executing the bootstrap may result in some bimodal realisations. However, it turnt out that this kind of situations happen rarely, especially the latter, so no ambiguity is encountered.

\section{Results}\label{sect3}

\begin{table*}
\small
\caption{Parameters of the fits. Label corresponds to panels in Fig.~\ref{fig1}. Errors are estimated using the bootstrap method.\label{tbl1}}
\begin{tabular}{c c c c c c c c c c c c}
\tableline
\tableline
Label & Dataset & $N$ & $i$ & $\mu_i$ & $\delta\mu_i$ & $\sigma_i$ & $\delta\sigma_i$ & $A_i$ & $\delta A_i$ & min. & $\delta$min. \\
\tableline
\multirow{2}{*}{(a)} & \multirow{2}{*}{BATSE 1B} & \multirow{2}{*}{226} & 1 & $-0.393$ & 0.099 & 0.465 & 0.069 & 0.272 & \multirow{2}{*}{0.040} & \multirow{2}{*}{2.158} & \multirow{2}{*}{0.049} \\
                     &                           & & 2 &  1.460 & 0.056 & 0.532 & 0.044 & 0.728 &       &      &   \\
\tableline
\multirow{2}{*}{(b)} & BATSE & \multirow{2}{*}{2041} & 1 & $-0.095$ & 0.051 & 0.627 & 0.033 & 0.336 & \multirow{2}{*}{0.018} & \multirow{2}{*}{3.378} & \multirow{2}{*}{0.272} \\
                     & current & & 2 &  1.544 & 0.018 & 0.429 & 0.013 & 0.664 &       &      &  \\
\tableline
\multirow{2}{*}{(c)} & \multirow{2}{*}{\it Swift} & \multirow{2}{*}{914} & 1 & $-0.026$ & 0.255 & 0.740 & 0.120 & 0.139 & \multirow{2}{*}{0.042} & \multirow{2}{*}{---} & \multirow{2}{*}{---} \\
                     &                            & & 2 & 1.638 & 0.031 & 0.528 & 0.023 & 0.861 &  &     &     \\
\tableline
\multirow{2}{*}{(d)} & \multirow{2}{*}{{\it BeppoSAX}} & \multirow{2}{*}{1003} & 1 & 0.626 & 0.186 & 0.669 & 0.075 & 0.355 & \multirow{2}{*}{0.084} & \multirow{2}{*}{---} & \multirow{2}{*}{---} \\
                     &                           & & 2 & 1.449 & 0.035 & 0.393 & 0.027 & 0.645 &       &     &     \\
\tableline
\multirow{2}{*}{(e)} & \multirow{2}{*}{\it Fermi} & \multirow{2}{*}{1596} & 1 & $-0.072$ & 0.073 & 0.525 & 0.044 & 0.215 & \multirow{2}{*}{0.021} & \multirow{2}{*}{2.049} & \multirow{2}{*}{0.248} \\
                     &                            & & 2 &  1.451 & 0.021 & 0.463 & 0.014 & 0.785 &                    &       & \\
\tableline
\end{tabular}
\end{table*}

\begin{figure*}
\includegraphics[width=0.9\textwidth]{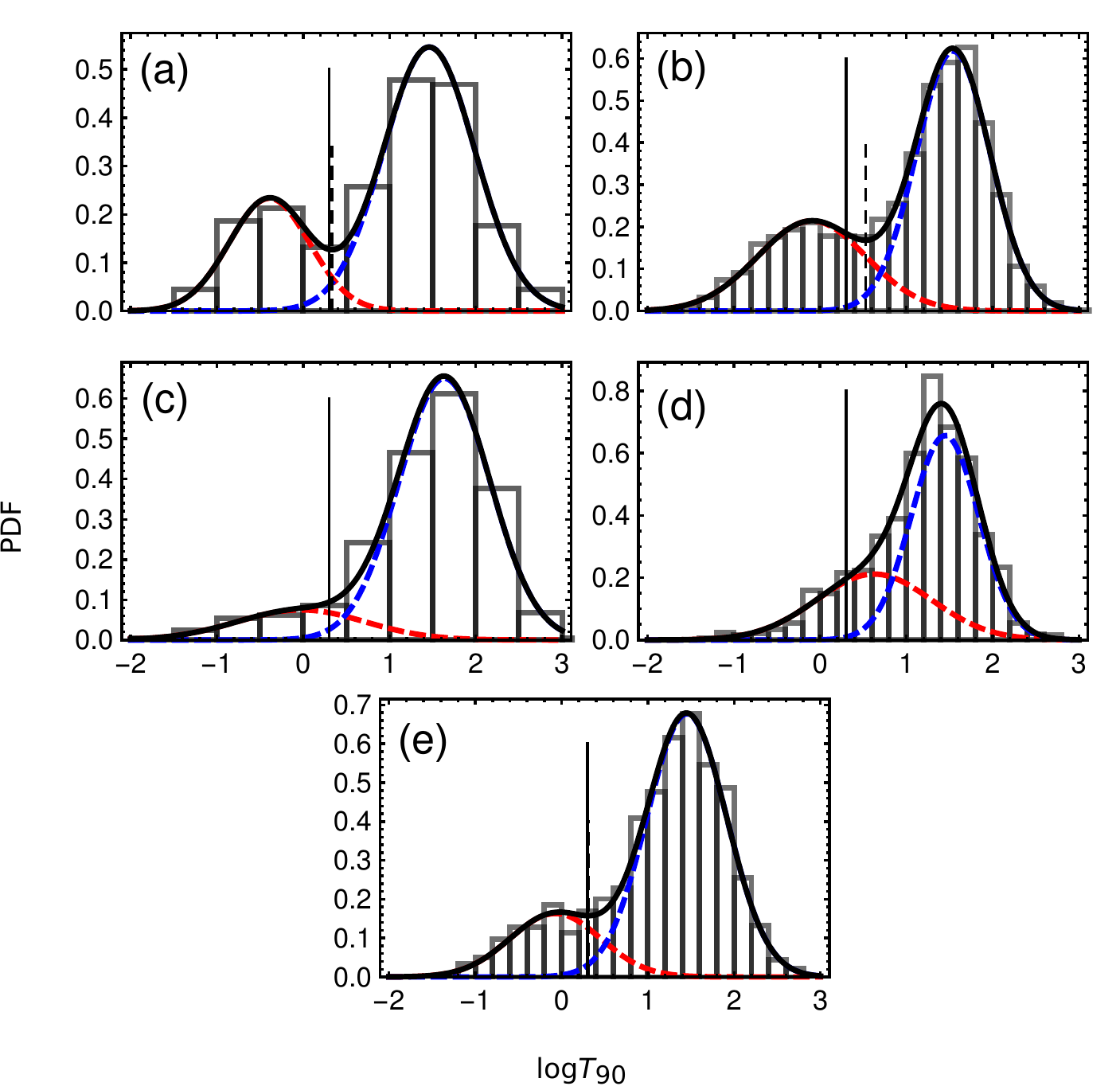}
\caption{Two-Gaussian PDFs fitted to $\log T_{90}$ data. Color dashed curves are the components of the (solid black) mixture distribution. Vertical solid line marks the conventional $2\,{\rm s}$ limit between short and long GRBs. Vertical dashed line marks the position of the local minimum (if present) of the mixture. The panels correspond to (a) BATSE 1B, (b) BATSE current, (c) {\it Swift}, (d) {\it BeppoSAX}, and (e) {\it Fermi} catalogs. In the latter, the new limit is very close to the conventional limit of \citet{kouve}.
\label{fig1}
}
\end{figure*}

The results in graphical form are displayed in Fig.~\ref{fig1}, where the vertical solid line marks the conventional limit of $2\,{\rm s}$, and the vertical dashed line marks the location of the minimum of a mixture of two normal distributions (a two-Gaussian). The {\it Swift} and {\it BeppoSAX} distributions are unimodal, so no new limit may be inferred. Among the 1000 bootstrap executions, 163 realisations were bimodal for the {\it Swift} sample, and only one yielded a local minimum in the case of {\it BeppoSAX}. Parameters of the fits are gathered in Table~\ref{tbl1}.

BATSE 1B yielded a minimum at $2.16\pm 0.05\,{\rm s}$, close to the value attained by \citet{kouve}. However, for the current BATSE catalog a limit of $3.38\pm 0.27\,{\rm s}$ is more suitable, and the conventional value of $2\,{\rm s}$ lies outside the interval more than five times of the error. {\it Fermi} dataset is the most consistent with the $2\,{\rm s}$ limit, yielding a minimum at $2.05\pm 0.25\,{\rm s}$. The shallower the minimum, the bigger the error obtained.

In the case of BATSE current catalog, all 1000 bootstrap realisations were bimodal; in BATSE 1B a minimum was present in 911 cases, while for {\it Fermi} there were 896 bimodal realisations.

\section{Discussion}\label{sect4}

\citet{bromberg} found, by constructing a PDF for collapsars and non-collapsars (a classification based on physical origin of a GRB), that while the $2\,{\rm s}$ criterion is suitable for BATSE, a value of $0.8\,{\rm s}$ is more appropriate for the {\it Swift} dataset. Herein, the $\log T_{90}$ distributions examined imply that the suitable limit between short and long GRBs for the BATSE current catalog should be $3.38\,{\rm s}$, based on a univariate analysis. This is significantly higher than the commonly applied $2\,{\rm s}$ criterion. In case of {\it Swift}, the duration distribution turnt out to be unimodal, and as such no natural limit may be inferred. Also {\it BeppoSAX} durations are unimodal, giving no new limiting value. It is important to note that the locations $\mu_1$, corresponding to the shorter component, are negative (hence $T_{90}<1\,{\rm s}$) for BATSE (1B and current), {\it Swift} and {\it Fermi}, while for {\it BeppoSAX} it is $\mu_1=0.626$, corresponding to $T_{90}\approx 4.23\,{\rm s}$. This is definitely not a short GRB group, and it is consistent with \citep{horvath09} where the short GRB group was not detected. The {\it Fermi} data have a minimum at $2.05\,{\rm s}$, consistent with the common limiting value.

The newly obtained limits result in different populations of short and long GRBs in the datasets examined (see Table~\ref{tbl2}). In BATSE 1B the fraction of long GRBs in the sample is unchanged (due to smallness of the sample; it appears there are no GRBs with durations between $2\,{\rm s}$ and $2.16\,{\rm s}$). In {\it Fermi} this fraction is nearly the same, slightly smaller than conventional due to the limit being slightly higher than $2\,{\rm s}$. The biggest difference is visible in the BATSE current catalog, where the new limit leads to diminishing the long GRBs fraction by 4\%.

\begin{table}
\small
\caption{Fractions of long GRBs and overlap of components of the two-Gaussian fits.\label{tbl2}}
\begin{tabular}{c c c c c}
\tableline
\tableline
\multirow{2}{*}{Label} & \multirow{2}{*}{Dataset} & \multicolumn{2}{c}{Long GRBs fraction [\%]} & \multirow{2}{*}{Overlap [\%]} \\
\cline{3-4}
      &         & Conventional\tablenotemark{a} & This work\tablenotemark{b} & \\
\tableline
(a) & BATSE 1B      & 73.89 & 73.89 & 5.68 \\
(b) & BATSE current & 75.50 & 71.53 & 10.1 \\
(c) & {\it Swift}   & 90.81 & ---   & 9.30 \\
(d) & {\it BeppoSAX}      & 88.24 & ---   & 34.4 \\
(e) & {\it Fermi}   & 83.40 & 83.08 & 9.06 \\
\tableline
\tablenotetext{a}{When the conventional \citep{kouve} limit of $2\,{\rm s}$ is applied.}
\tablenotetext{b}{When the new limits obtained herein are applied.}
\end{tabular}
\end{table}

All catalogs are dominated by long GRBs, the highest fraction of more than 90\% being observed by {\it Swift}. In BATSE current, the proportion of short and long GRBs is $\sim 1:3$. {\it Swift} is more sensitive in soft bands (corresponding to long GRBs) than BATSE was, while {\it Fermi}'s sensitivity at very soft and very hard GRBs had increased compared to BATSE \citep{meegan09}. {\it BeppoSAX} is also more sensitive to long GRBs due to the trigger system which used $1\,{\rm s}$ as short integration time \citep{horvath09}, hence the lack of a distinct short GRB peak. GRBs tend to be softer at later times, hence the inferred duration is shorter than it might be. This naturally leads to a conclusion that the duration distributions as observed by different satellites must differ between each other, and also the limit between short and long GRBs (the local minimum) has to be placed at different locations. 

The duration $T_{90}$ itself is not an unambiguous indicator of a GRB type, as the components of the fitted two-Gaussians overlap strongly. To quantify this overlap, the common area under the curves is computed (the total area of a two-Gaussian PDF is equal to unity, and the area under each component is given by the weights $A_i$). This gives a probability of misclassifying a GRB from 5.68\% (BATSE 1B) to 10.1\% (BATSE current), and an enormous 34.4\% in the case of {\it BeppoSAX} (see Table~\ref{tbl2}).

A solution, proposed to deal with the classification ambiguity problem, was proposed and examined in a number of papers \citep{hakkila,horvath04,horvath06,chatto,veres,horvath10}. The idea is to examine a multi-dimensional space of various parameters; particularly, a two-dimensional space of the hardness ratio vs. duration $T_{90}$. This approach still awaits to be applied to the {\it Fermi} GRBs. Additional parameters have been defined and proposed for GRB classification as well. Examples are \mbox{$\varepsilon=E_{\gamma,{\rm iso},52}/E_{p,z,2}^{5/3}$} (unambiguously dividing short and long GRBs) \citep{lu}, minimum variability time-scale (MVTS) \citep{bhat,maclach12,maclach13a,maclach13b,golkhou,golkhou2} or Hurst exponent (HE) \citep{maclach13b,Tarnopolski3}. Still, the most common criterion is the GRB duration, and its limitting value has been shown herein to be detector dependent.

\section{Conclusions}\label{sect5}

The duration distributions of various catalogs (BATSE 1B, BATSE current, {\it Swift}, {\it BeppoSAX},  and {\it Fermi}) were examined. A mixture of two Gaussians was fitted to the $\log T_{90}$ distributions in search for a new limiting value placed at the local minimum. It was found that the datasets from {\it Swift} and {\it BeppoSAX} are unimodal, hence no new limit may be inferred. The results from BATSE 1B and {\it Fermi} are consistent with the conventional phenomenological limit of $2\,{\rm s}$ \citep{kouve}, whereas in BATSE current catalog the value obtained is equal to $3.38\pm 0.27\,{\rm s}$. This leads to a different than commonly established, fraction of long GRBs in the sample, diminished by 4\% (see Table~\ref{tbl2}).

Due to the significant overlap and dependence of the location of the minimum on the detector, while the division into short and long GRBs based on their durations is qualitatively proper, it is not unambiguously related to its progenitor, i.e. collapsar or non-collapsar. Therefore, as the short-long phenomenological classification justifies the existence of two distinct GRB classes, it gives limited insight into the underlying physical phenomenon.

\end{document}